\documentclass[superscriptaddress,amsmath,amssymb]{revtex4}
\usepackage{graphicx}
\usepackage{dcolumn}
\usepackage{bm}
\usepackage{ulem}
\usepackage{xfrac}
\usepackage{lipsum}
\usepackage{natbib}

\usepackage[printwatermark]{xwatermark}
\usepackage{xcolor}

\usepackage{graphicx,amsmath,amssymb}
\usepackage{epstopdf}
\usepackage{lineno}
\usepackage[colorlinks=true, urlcolor=blue, linkcolor=blue, citecolor=blue]{hyperref}

\begin{document}

\title{Hubble Tension and Gravitational Self-Interaction}

\author{Corey Sargent}
\affiliation{Department of Physics, Old Dominion University,
Norfolk, Virginia 23529, USA}

\author{Alexandre Deur}
\affiliation{Department of Physics, University of Virginia,
Charlottesville, Virginia 22901, USA}
\affiliation{Department of Physics, Old Dominion University,
Norfolk, Virginia 23529, USA}

\author{Bal{\v s}a Terzi{\'c}}
\affiliation{Department of Physics, Old Dominion University,
Norfolk, Virginia 23529, USA}


\begin{abstract}
One of the most important problems vexing the $\Lambda$CDM 
cosmological model is the Hubble tension. 
It arises from the fact that 
measurements of the present value of the
Hubble parameter performed with low-redshift quantities, e.g., the Type IA supernova, tend to
yield larger values than  measurements from quantities originating at high-redshift, e.g., fits of cosmic microwave background radiation.
It is becoming likely
that the discrepancy, currently standing at $5\sigma$,  is not due to  systematic errors in 
the  measurements. 
Here we
 explore whether the self-interaction of gravitational fields in General Relativity, which are traditionally neglected
when studying the evolution of the universe, can explain the tension.
We find that with  field self-interaction accounted for, both low- and high-redshift  data are
{\it simultaneously} well-fitted, thereby showing that gravitational self-interaction could explain the Hubble tension.
 Crucially,  this is achieved without introducing additional parameters.
\end{abstract}




\maketitle

\section{The Hubble Tension \label{sec:Hubble-tension}}
Modern cosmology began with the discovery of Hubble's law. Its central element, the present value of the Hubble parameter, $H_0$, has a troubled history of measurements and it is only in the last two decades since precise determinations became available. 
However, two types of precision measurements of $H_0$ are in conflict. The first type comprises observations of phenomena originating at high redshift $z$, principally the  power spectrum of the cosmic microwave background (CMB)~\citep{Planck:2018vyg} and  the baryon acoustic oscillations (BAO)~\citep{eBOSS:2020yzd}.
The second type consists of determination of $H_0$ from  low-$z$ phenomena, notably  using standard candles~\citep{Riess:2016jrr} and time-delay cosmography~\citep{Wong:2019kwg} methods. See~\cite{Abdalla:2022yfr} for the low- and high-$z$ methods providing $H_0$. The high-$z$ phenomena yield $H_0$ values significantly lower than those from low-$z$. This is known as the ``Hubble tension''~\citep{Verde:2019ivm, DiValentino:2021izs, Abdalla:2022yfr}.
The discrepancy presently reaches a {{5}}$\sigma$ significance:
the combined high-$z$ measurements yield $67.28 \pm 0.60$~km/s/Mpc
while the combined low-$z$ measurements yield {{$H_0=73.04 \pm 1.04$}}~km/s/Mpc \citep{Riess_2022}. Yet, individual low-$z$ measurements can be as much as 6$\sigma$ away~\citep{Abdalla:2022yfr} from the most precise high-$z$ datum, the Planck satellite result~\citep{Planck:2018vyg}.

Although the Hubble tension may originate from unaccounted systematic effects {\citep{Freedman_2020}}, the consistency of the high-$z$ results on the one hand, and that of the low-$z$ results on the other, suggests that it could instead reveal a limitation
of the current standard model of cosmology, the dark energy-cold dark matter model ($\Lambda$CDM). 
This would be just one of the several malaises of $\Lambda$CDM. A first worry is that detection of dark matter particles by direct~\citep{Kahlhoefer:2017dnp} or 
indirect~\citep{Gaskins:2016cha} measurements is still wanting,
with searches having almost exhausted the allowed parameter spaces of likely candidates.
Furthermore, the most natural extensions of the standard model of particle physics which offer convincing dark matter candidates are mostly ruled out, e.g., 
minimal SUSY~\citep{Arcadi:2017kky}. Other worries with 
$\Lambda$CDM include overestimating the number of globular clusters and dwarf galaxies~\citep{Klypin:1999uc}
or the lack of uncontrived explanation for tight correlations between the supposedly sub-dominant baryonic matter and quantities characterizing galaxy dynamics, e.g.,   
the Tully-Fisher relation~\citep{Tully:1977fu},  radial acceleration relation (RAR)~\citep{McGaugh:2016leg}, or Renzo's rule~\citep{renzo}. 
These issues motivate developing alternatives to 
$\Lambda$CDM that could naturally resolve these problems. 
Here we follow this direction and investigate whether the Hubble tension can be  understood
with a model that incorporates the fact that in General Relativity (GR), gravitational fields interact with each others (field self-interaction, SI).
That central feature of GR is the basis for the GR-SI model. This model already explains the chief observations involving dark matter/energy without recourse to dark components:  
the flat rotation curves of galaxies~\citep{Deur:2009ya, Deur:2020wlg}; 
the high-$z$ supernova luminosities~\citep{Deur:2017aas};
the CMB anisotropies~\citep{Deur:2022ooc}; 
the  formation of large structures~\citep{Deur:2021ink};
the matter power spectrum~\citep{Deur:2022ooc};
the internal dynamics of galaxy clusters, including that of the  Bullet Cluster~\citep{Deur:2009ya}; and the  RAR~\citep{Deur:2019kqi} and Tully-Fisher~\citep{Deur:2009ya} relations.

In the next section, we recall the physical basis
of the GR-SI framework and its predictions.  We then discuss  how, from the perspective of the GR-SI model, a Hubble tension should arise if low- and high-$z$
data are analyzed with $\Lambda$CDM, and why the
tension is not present in GR-SI. After summarizing
how the evolution of the universe affects the CMB anisotropy observations in both the 
GR-SI and $\Lambda$CDM frameworks, we use GR-SI to fit luminosity distance data. This constrains the GR-SI parameters describing the effects of large-scale structure formation on the long distance propagation of gravity, effects that are encapsulated in a so-called {\it depletion function} $D_M(z)$. 
Finally, we verify that with the constrained parameters, { the GR-SI fit} reproduces { better} the CMB power spectrum with the low-$z$ value of $H_0$ { than with the high-${z~H_0}$ determination. 
We also find that if ${H_0}$ is left a free parameter, its best fit value agrees with the low-${z}$ determination rather than the high-${z}$ one. This indicates an} absence of Hubble tension in the GR-SI model.
We will consider only the scalar multipole coefficient ${ C^s_{TT,l} }$ since it is sufficient to investigate whether a Hubble tension is present in the GR-SI model. In particular, it is not necessary for the goal of this article to investigate the polarized CMB data.
 
\section{Field self-interaction and its consequences \label{sec:FSI}} 

A defining feature of GR is that it is a non-linear theory: gravity fields interact with each other, in contrast
to Newtonian gravity. The linear character of the latter allows for  the field superposition principle, 
while in GR, the combination of fields differ from their sum since the fields interact.
In fact, the GR Lagrangian
$\mathcal{L}_{\mathrm{GR}}={\sqrt{\det(g_{\mu\nu})}\, g_{\mu\nu}R^{\mu\nu}}/{(16\pi G)}$
(here $g_{\mu\nu}$ is the metric, $G$  is Newton's constant and $R_{\mu\nu}$ is the Ricci tensor) expressed in a polynomial form~\citep{Zee}:
\begin{eqnarray}
\mathcal{L}_{\mathrm{GR}}\!=\! \sum_{n=0}^\infty (16\pi MG)^{n/2}\left[\phi^n\partial\phi\partial\phi\right], 
\label{eq:Polynomial Einstein-Hilber Lagrangian}
\end{eqnarray}
explicitly shows 
that a gravitational field self-interacts. Here, $\phi_{\mu\nu}$ is
the gravitational field due to a unit mass
and is defined as the deviation of $g_{\mu\nu}$ 
from a reference constant metric $\eta_{\mu\nu}$,  $\phi_{\mu\nu} \equiv (g_{\mu\nu} - \eta_{\mu\nu})/\sqrt{M} $, 
where $M$ is the mass of the system. 
For simplicity, we ignored the matter term of $\mathcal{L}_{\mathrm{GR}}$: to discuss the pure field case is sufficient.
The bracket in $\left[\phi^{n}\partial\phi\partial\phi\right]$ signifies a sum of Lorentz-invariant terms whose forms are
$\phi^{n}\partial\phi\partial\phi$, e.g., $\left[\partial\phi\partial\phi\right]$ is the Fierz-Pauli Lagrangian of linearized GR~\citep{Fierz:1939ix}.
Newtonian gravity is recovered  if  $\eta_{\mu\nu}$ is the Minkowski metric and if one keeps only
the time-time component of the $n=0$ term of Eq.~(\ref{eq:Polynomial Einstein-Hilber Lagrangian}): 
$\left[\partial\phi\partial\phi\right] \partial^{\mu}\phi_{00}\partial_{\mu}\phi^{00}$ 
and $\partial^0\phi_{00}=0$. The term $\left[\partial\phi\partial\phi\right]$ formalizes the free motion of the field, {\it viz}, it generates the 
two-point correlation function that gives the probability for the field to freely propagate from one spacetime point to another. The $n>0$ terms are interaction terms and therefore 
cause the field SI.
An analogous phenomenon occurs for the nuclear Strong Force, whose theory is 
Quantum Chromodynamics  (QCD). Actually, the reason why GR and QCD are non-linear theories is the same: they possess several types of distinct ``charges''. For GR, they are the mass/energy, momentum and stress. For QCD, they are the three color charges. This causes the fields  of GR and QCD to be rank-2 tensors, i.e., non-commuting objects. The non-zero
commutators in turn give rise to SI terms.  It results in GR and QCD having the same Lagrangian structure. Field SI is a central and conspicuous feature of QCD due to its large coupling $\alpha_s$~\citep{Deur:2016tte}.
In contrast, field SI in GR is controlled by $\sim$$\sqrt{GM/L}$ (with $L$ a characteristic length 
of the system), whose value is typically small. This makes the linear approximations of GR, 
e.g.,  the Newtonian or the Fierz-Pauli theories, adequate for most applications. 
However, if $\sqrt{GM/L}$ is large enough, SI {\it must} be accounted for: it is an unavoidable consequence of GR. The calculations in~\cite{Deur:2009ya, Deur_DM-EPJC, Deur:2020wlg} indicate that 
for galaxies or galaxy clusters, $\sqrt{GM/L}$ is large enough to enable 
SI. 

One consequence of SI in QCD is to enhance the binding of quarks, resulting in 
their confinement. 
Likewise in GR, if a galactic mass is large enough to enable SI, it would enhance the binding of galactic components in a manner that directly 
leads to flat galactic rotation curves~\citep{Deur:2009ya} without requiring dark matter. 
The increased binding also dispenses with the need for dark matter to account for the growth of 
large-scale structures~\citep{Deur:2021ink}. 
On the other hand, using Newtonian gravity to analyze systems in which SI is important
overlooks the binding enhancement and produces an apparent mass discrepancy
interpreted as dark matter.
Importantly, SI effects cancel out in isotropic and homogeneous systems. For example,
a nearly spherical galaxy has much less evidence of dark matter than a flatter
galaxy~\citep{Deur:2013baa, Winters:2022tew}.
 
Another direct and crucial consequence of the  binding enhancement comes from energy conservation: the
increase of binding energy inside a system must be balanced by a reduction of the gravitational energy outside of the system. 
In QCD, the larger binding confines quarks into  hadrons, while outside hadrons, the Strong Force 
declines into the  much weaker residual Yukawa interaction.
Likewise, if SI binds more tightly massive systems, gravitation must be reduced outside these systems. 
Overlooking that large-distance reduction of gravity would require a compensating global repulsion in much the same way as overlooking the binding enhancement requires a compensating dark mass. The purported repulsion would then be interpreted as dark energy. 

The enhanced binding of structures, {\it viz}, the {\it local} effect of SI,
is computed starting from GR's Lagrangian, Eq.~(\ref{eq:Polynomial Einstein-Hilber Lagrangian})~\citep{Deur:2009ya, Deur_DM-EPJC}.
The large-distance suppression of gravity, {\it viz}, the {\it global} effect, is evaluated effectively
using a {\it depletion function} $D_M (z)$ that originates from
lifting the traditional assumptions that the universe is isotropic and homogeneous~\citep{Deur:2017aas}. 

\begin{figure}
\center
\includegraphics[width=0.49\textwidth]{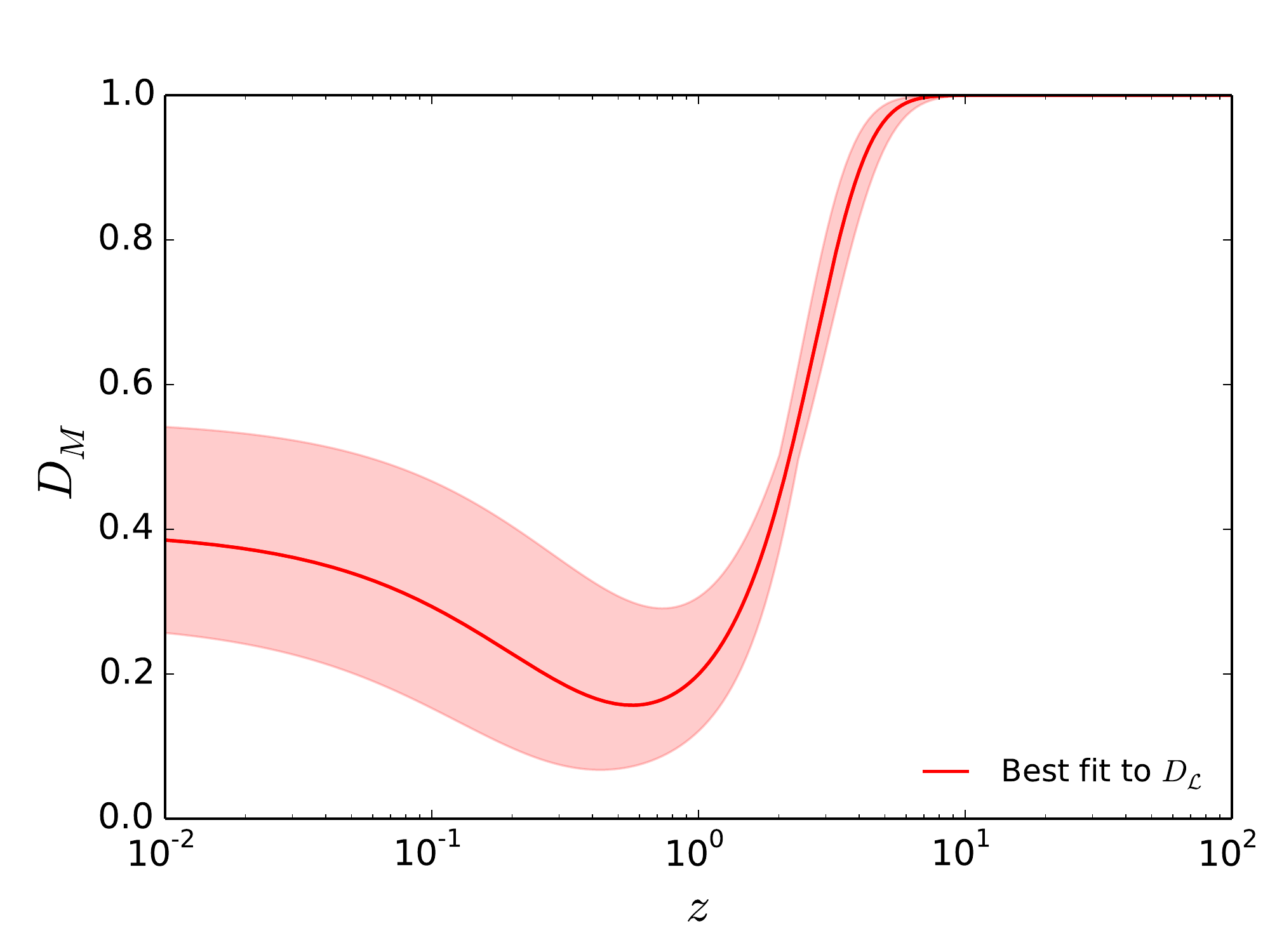}
\vspace{-0.4cm}
\caption{Depletion function $D_{M}(z)$ determined from the optimizing the fit to the low- and high-$z$ $D_{\mathcal L}$ data in Fig.~\ref{fig:DL(z)}.}
\label{fig:D(z)}
\end{figure}

If $D_M=0$, gravity is fully quenched at large-distance while for $D_M=1$
there is no net SI effect. Thus, $D_M(z) \approx 1$ for the early universe since it was 
nearly isotropic and homogeneous. In contrast, the large-scale structures of the present universe entail
$D_M(z\approx0) < 1$. The form of $D_M(z)$ first proposed in~\citep{Deur:2017aas} can be approximated by:
\vspace{-0.1cm}
\begin{equation}
D_M(z)= 1-(1+e^{(z-z_0)/\tau})^{-1}+Ae^{-z/b}.
\label{eq:Dep_nom}
\end{equation} 
Here, $z_0$ is the redshift characterizing the large-scale structure formation epoch and  $\tau$ its duration. 
$A$ is the mass fraction of structures whose shapes have evolved into more symmetric ones (e.g., disk galaxies merging to form
elliptical galaxies) and $b$ is the duration of that evolution process. 
Fig.~\ref{fig:D(z)} displays $D_M(z)$. 

\section{The Hubble Tension from the GR-SI perspective}

A Hubble tension arising within $\Lambda$CDM is expected from the perspective of GR-SI: $H_0$ affects the observation of the CMB anisotropies 
essentially {\it via} the angular diameter distance of last scattering, $d_A$. 
This quantity depends upon the evolution of the universe
 similarly to the luminosity distance $D_{\mathcal L}$ 
that enters the lower-$z$ determination of $H_0$, e.g., {\it via} supernova
observations. Specifically, $d_A(z)=D_{\mathcal L}(z)/(1+z)^2$.
For example, in the $\Lambda$CDM model,
\begin{eqnarray}
  d_A(z)&=& \frac{1}{H_0(1+z)\sqrt{\Omega_K}}\sinh\left(\sqrt{\Omega_K}\int^1_{(1+z)^{-1}} \frac{dx}{\sqrt{\Omega_\Lambda x^4+\Omega_K x^2 +\Omega_M x +\Omega_\gamma}}\right), \\
  D_{\mathcal L}(z)&=& \frac{(1+z)}{H_0\sqrt{\Omega_K}}\sinh\left(\sqrt{\Omega_K}\int^1_{(1+z)^{-1}} \frac{dx}{\sqrt{\Omega_\Lambda x^4+\Omega_K x^2 +\Omega_M x +\Omega_\gamma}}\right)
  , 
\end{eqnarray}
with $\Omega_\Lambda$, $\Omega_M$ and $\Omega_\gamma$ the dark energy, total matter and radiation densities relative to the critical density, respectively, and $\Omega_K\equiv \sfrac{K}{a_0^2 H_0^2}$ with $K$ the 
curvature and $a_0$ the  Friedmann-Lema\^itre-Robertson-Walker scale factor at present time.
Therefore, the determination of $H_0$ from CMB observations is analogous to a highly accurate $D_{\mathcal L}(z_L)$ 
observation, where $z_L$ is the redshift  at the time of last rescattering. 

\begin{figure}
\center
\includegraphics[width=0.497\textwidth]{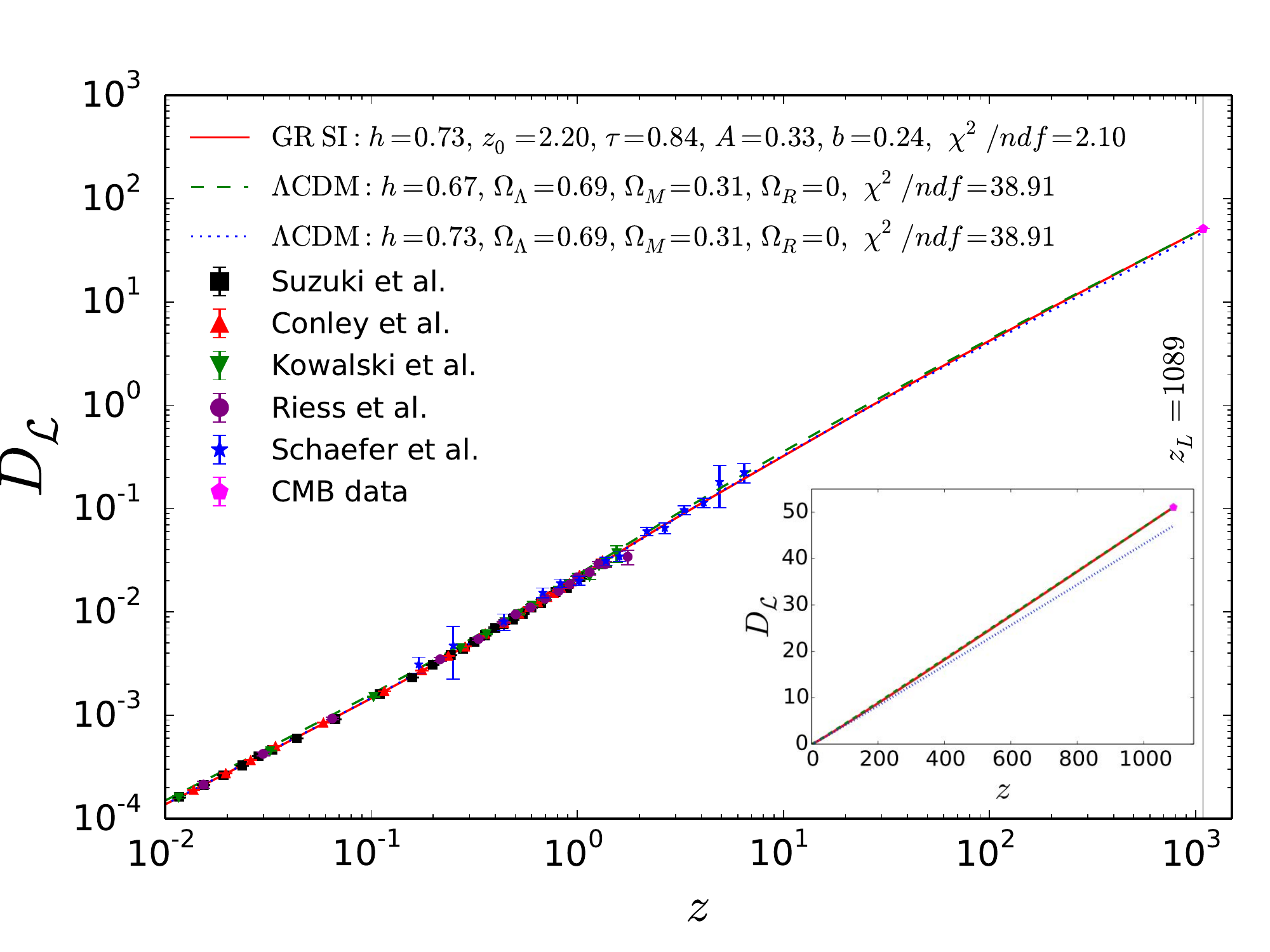}
\includegraphics[width=0.497\textwidth]{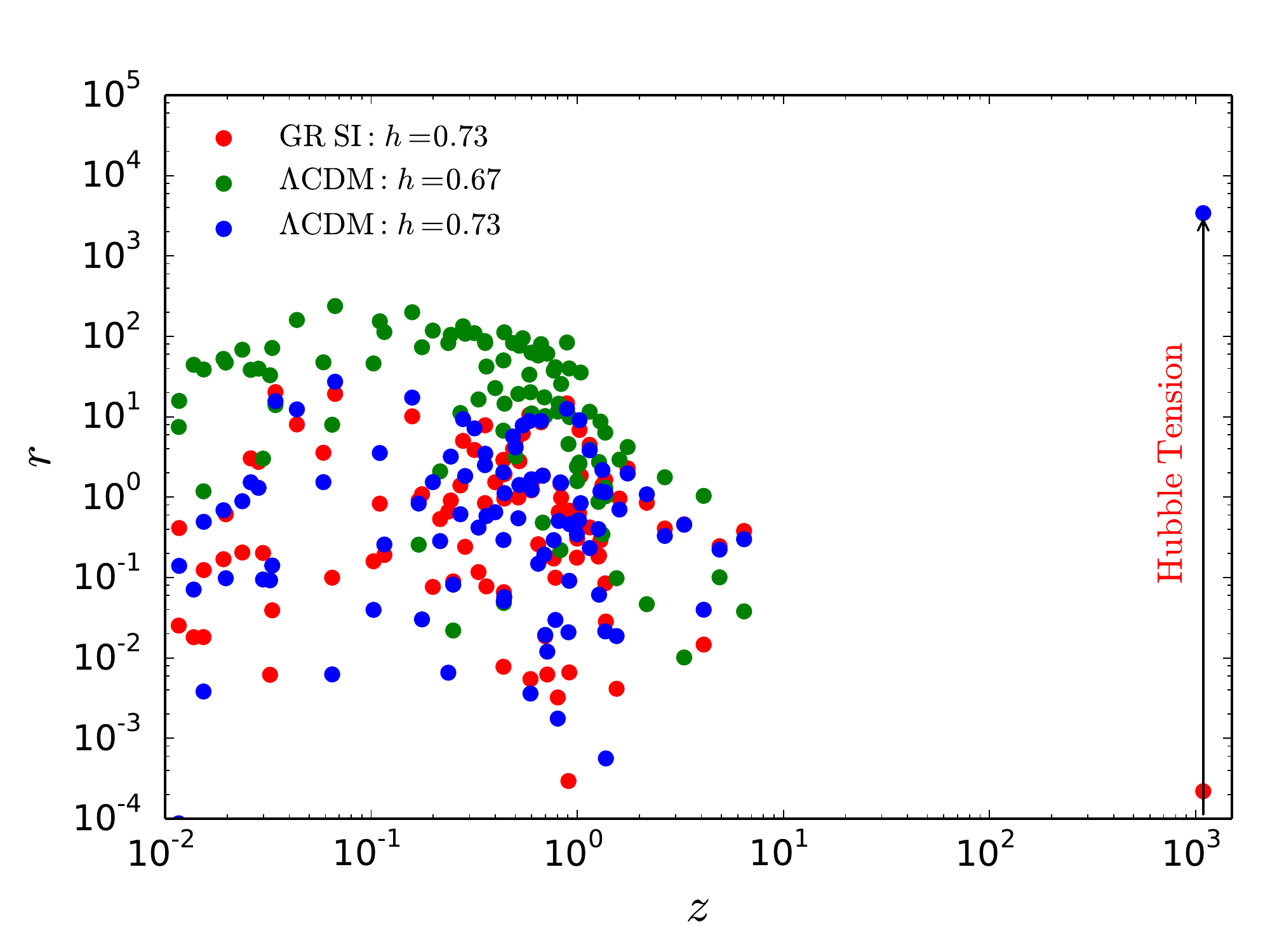}
\vspace{-0.3cm}
\caption{Left: Luminosity distance $D_{\mathcal L}$ as a function of redshift
$z$ for: $\Lambda$CDM using $h\equiv \sfrac{H_0}{100 \rm{~km/s/Mpc}}=0.67$ (dashed green line) or $h=0.73$ (dotted blue line); 
and GR-SI with $h=0.73$ (solid red line). 
The embedded figure is the same but in linear rather than log scales. 
The low-$z$ observational data, shown by the square, triangle, circle and star symbols, are normalized using the $h=0.73$ average low-$z$ determination.
The pentagon symbol shows $D_{\mathcal L}(z_L)$ as it would be obtained using the values of $z_L$ and $H_0$ from the $\Lambda$CDM fit of the CMB. 
Right: Same as the left panel but for the normalized residual $r=(D_{\mathcal L}-d_{\rm obs})^2/e_{\rm obs}^2$, where $d_{\rm obs}$ is the observed data, $e_{\rm obs}$ their uncertainty, and the colors match that of the three different models used to compute $D_{\mathcal L}$ in the left panel. The Hubble tension appears as the offset between the $\Lambda$CDM curve which fits the low-$z$ data (dotted blue line in the left panel and blue dots in the right panel) and the blue dot at $z_L$.
The green dot at $z_L$ is near $r=0$ and hence not visible with the log scale.
}
\label{fig:DL(z)}
\end{figure}

Figure \ref{fig:DL(z)} depicts  two luminosity  distances 
$D_{\mathcal L}(z)$ 
calculated within $\Lambda$CDM
with $\Omega_{\Lambda}=0.69$, $\Omega_{M}=0.31$ and $K=0$, but
different  $H_0$ values: $73.06$~km/s/Mpc, which matches the
supernova and $\gamma$-ray data at low-$z$ (dashed blue line in the left
panel and blue dots in the right), and the other with 
$67.28$~km/s/Mpc to match the CMB $D_{\mathcal L}(z_L)$ 
(dotted green line and green points).
The uncertainty of the CMB  datum is adjusted to equalize the $\chi^2/ndf$ values of the fits for the comparison of the data and the two $\Lambda$CDM cosmologies. 
The Hubble tension is evident in 
the two $\Lambda$CDM curves which match well either the low-$z$ data or the
high-$z$ data, but not both. However, the GR-SI model for $D_{\mathcal L}(z)$~\citep{Deur:2017aas},
\begin{eqnarray}
  D_{\mathcal L}(z)= \frac{(1+z)}{\sqrt{\Omega_K}H_0}\sinh\left(\sqrt{\Omega_K} \int^1_{1/(1+z_L)} \frac{dx}{\sqrt{\Omega_K x^2 +D_M(1/x-1) x}}\right), 
  \label{eq:DL-GR-SI}
\end{eqnarray} 
fits both data sets well, as quantified by a significantly smaller 
$\chi^2/ndf$ value, 
therefore exhibiting no signs of Hubble
tension.
Here, we elected to let the parameters of $D_M(z)$ be determined from the best fit to the $D_{\mathcal L}(z)$ data. This yields $z_0=2.20\pm0.18$, $\tau=0.84^{+0.15}_{-0.19}$, $A=0.33\pm0.09$ and $b=0.24^{+0.10}_{-0.16}$. Originally the values of the
parameters were obtained from the knowledge of the evolution of large-scale structures.
The value $z_0=2.20\pm0.18$ is smaller than the estimate from large structure formation, $z_0=6.3^{+1.6}_{-2.0}$~\citep{Deur:2021ink}, but the  ratio $z_0/\tau=2.62$ happens to be the same for the fit and the estimates from large structure formation. The fit values for the $A$ and $b$ parameters agree with the earlier values, $A=0.25^{+0.20}_{-0.17}$ and $b=0.20^{+0.15}_{-0.05}$.

The $D_{\mathcal L}(z)$ calculated within $\Lambda$CDM and GR-SI differ chiefly at intermediate values of $z$ because SI induces a 
large-distance suppression of gravity which curves $D_{\mathcal L}(z)$ in the $1 \lesssim z \lesssim10$ domain, when large-scale structures 
 start forming~\citep{Deur:2017aas, Deur:2022ooc}. 

The specific timing and amount of matter involved in the formation of large-scale structures result in the particular $z$-dependence of $D_M(z)$ which differs from the $\propto z^4$ effect of dark energy in $\Lambda$CDM. Thus, if SI
noticeably influences the evolution of the universe, there will arise a discrepancy with $D_{\mathcal L}(z)$ determinations using smaller-$z$ phenomena for which the evolution spans a much smaller range. 
Since the determination of $H_0$ from the CMB is analogous to a determination using $D_{\mathcal L}(z_L)$, extracting $H_0$ from
the CMB using the $\Lambda$CDM framework will cause a tension with $H_0$ measurements at lower $z$.
The same applies to the baryonic acoustic oscillations (BAO) observation from the CMB. It is characterized by the acoustic horizon angular size, $\theta=\sfrac{d_H}{d_A(z_L)}$, where $d_H$ is the acoustic horizon. 
Since $d_H$ is the comoving distance travelled by a sound wave until recombination, {\it viz}, it happens for $z>z_L$ when the universe was homogeneous and dark energy negligible, $d_H$ is essentially the same for $\Lambda$CDM and GR-SI. It is the distinct evolution of $d_A(z)$ in $\Lambda$CDM and GR-SI that makes their $\theta$ predictions different. Like $D_{\mathcal L}$, $d_A$ is predicted by $\Lambda$CDM to be larger at $z=0$, yielding smaller $\theta$ and $H_0$ values compared to local measurements and the expectation from GR-SI.
 

\section{Dependence of the CMB observations on the Expansion of the Universe}

{ The GR-SI fit of ${ D_{\mathcal L}(z)}$ just discussed indicates
that there is no Hubble tension in the GR-SI model. An independent test that would support this conclusion is to fit the CMB within the GR-SI framework, and check 
that the low-${z ~H_0}$ determination provides a better fit to the CMB than the
high-${z ~H_0}$ one.}
We use an analytical expression of the CMB anisotropies to show how the expansion of the universe affects their observations at present-day. 
{ 
Such analytical expression is provided by the hydrodynamic approximation~\cite{Weinberg:2008zzc}. Despite not being as accurate as state-of-the-art numerical treatments of the CMB, this treatment is sufficient for the goal of this article, namely to investigate the Hubble tension within the GR-SI model. This is verified {\it a posteriori} by the small ${\chi^2/ndf}$ characterizing the GR-SI fits to the CMB.}
At $z_L$, the universe is very homogeneous, making SI effects negligible. Thus, the phenomena 
that created the CMB anisotropies are unaffected and so are the mathematical expressions formalizing them. However, some of the parameters 
entering the CMB anisotropy expression use their present time values. They are thus affected by the expansion of the universe
and therefore contribute to the Hubble tension. 
In what follows, values of parameters at the present time, matter-radiation equilibrium time, and last scattering time 
are indicated by the subscripts $0$, $EQ$ and $L$, respectively.
Baryon relative density is denoted by 
$\Omega_B$, and, for $\Lambda$CDM, the dark matter relative density is $\Omega_{DM}$.
We consider $C^s_{TT,l}$, the scalar multipole coefficient for 
the temperature-temperature angular correlation (here $l$ is the multipole moment). Its expression within the 
{\it hydrodynamic approximation} is provided in \cite{Weinberg:2008zzc}:
\begin{eqnarray}
 \frac{l(l+1)C^s_{TT,l}}{2\pi}&=&\frac{4\pi T_0^2 N^2 e^{-2\tau_{reion}}}{25}\int_1^\infty d\beta \bigg(\frac{\beta l}{l_\mathcal{R}}\bigg)^{n_s-1} 
\bigg\{ \frac{3\sqrt{\beta^2-1}}{\beta^4 (1+R_L)^{\sfrac{3}{2}}} \mathcal{S}^2(\beta l /l_T) e^{-\sfrac{2\beta^2 l^2}{l_D^2}}
\sin^2\big(\beta l/l_H+\Delta(\beta l /l_T) \big) + \nonumber \\
&&  \frac{1}{\beta^2\sqrt{\beta^2-1}}\bigg[3\mathcal{T}(\beta l /l_T)R_L - (1+R_L)^{\sfrac{-1}{4}} \mathcal{S}(\beta l /l_T) e^{-\sfrac{\beta^2 l^2}{l_D^2}}
\cos\big(\beta l/l_H+\Delta(\beta l /l_T) \big)\bigg]^2 \bigg\}+\mathcal{C}(l).
\label{eq:hydro_approx}
\end{eqnarray}
The first term in the curly bracket formalizes the Doppler effect.
The second term provides the Sachs-Wolf and intrinsic temperature anisotropy effects. Both terms also contains the large-$l$ damping. 
$N$ is the normalization of the primordial perturbations, 
$\tau_{reion}$ is the reionized plasma optical depth,
$\beta$ is an integration variable akin to a wave number, $n_s$ is the scalar spectral index, 
$l_\mathcal{R}\equiv(1+z_L)k_\mathcal{R}d_A$ is a multipole characteristic value, with  $k_\mathcal{R}\equiv0.05~$Mpc$^{-1}$ 
a conventional scale. 
Other multipole characteristic values are
$l_T=\sfrac{d_A}{d_T}$ ($d_T$ is a  length scale whose form differs in $\Lambda$CDM and GR-SI; see below),
$l_D=\sfrac{d_A}{d_D}$ ($d_D$ is the damping length) and
$l_H=\sfrac{d_A}{d_H}$.
$R_L=\sfrac{3\Omega_B}{4\Omega_\gamma(1+z_L)}$ is a ratio of relative densities and
$\mathcal{S}$, $\mathcal{T}$ and $\Delta$ are transfer functions. 
Finally, $\mathcal{C}(l)$ is a second-order term correcting the approximations of the hydrodynamic model~\citep{Deur:2022ooc}. Hereafter, since $\mathcal{C}(l)$ is small, we will ignore its possible dependence on the difference between the universe evolutions according to $\Lambda$CDM and GR-SI.
{ 
The integrated Sachs-Wolf, Sunyaev-Zel'dovich and
cosmic variance effects, which produce anisotropies that are extrinsic to the CMB origin, are not included in the hydrodynamic model. This does not affect our study of  the Hubble tension since we will focus on  the multipole range ${ 48<l<1800}$, a domain where these effects are unimportant.}

\begin{figure}
\center
\includegraphics[width=0.5\textwidth]{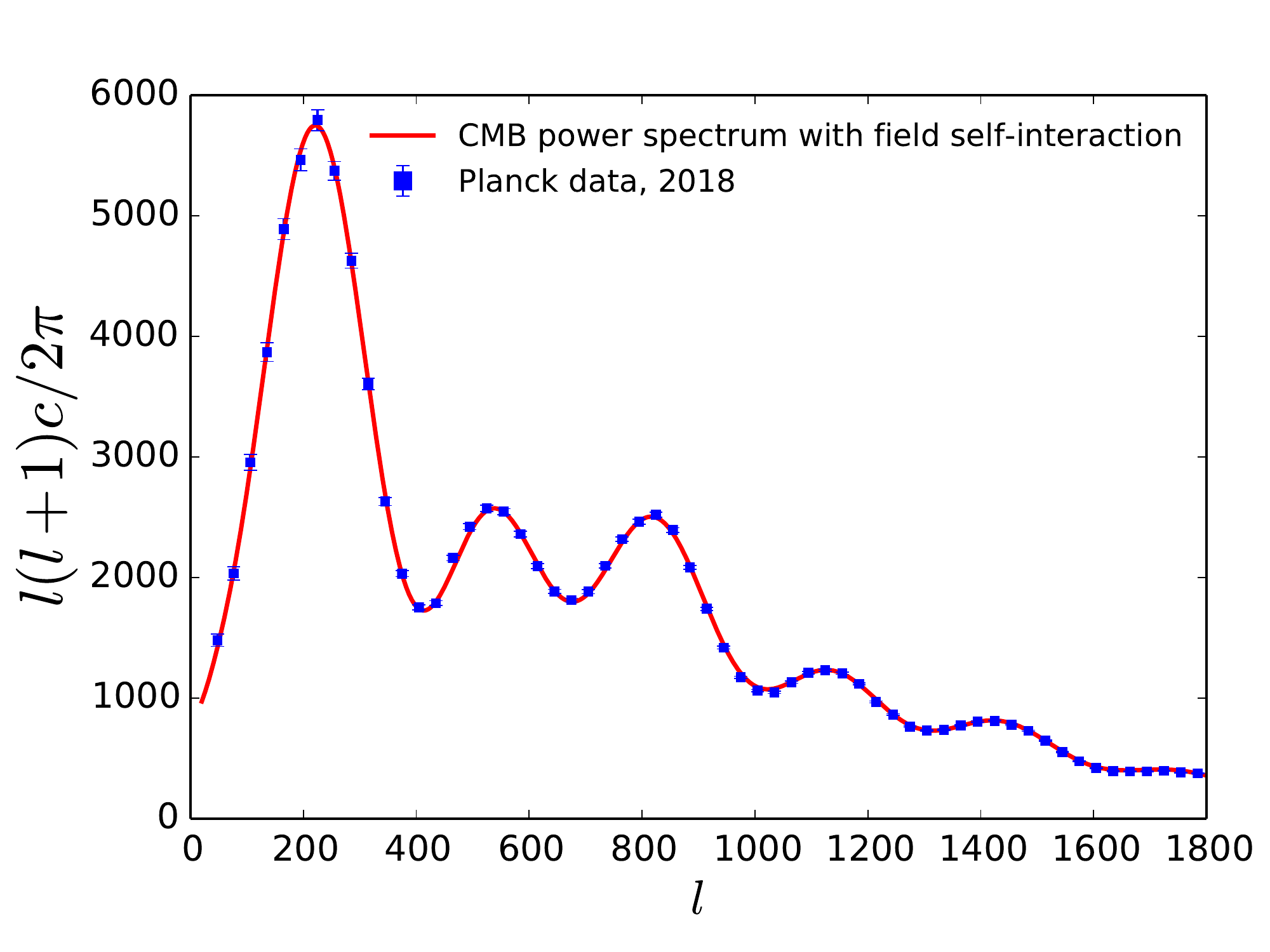}
\vspace{-0.3cm}
\caption{Power spectrum of the CMB temperature anisotropy. 
The continuous line is $l(l+1)C^s_{TT,l}/(2\pi)$ computed 
using GR-SI with the low-$z$ average for the Hubble parameter, $H_0=73.06$~km/s/Mpc. 
The squares are the Planck measurement (\cite{Aghanim:2019ame}, 2018 release).
}
\label{fig:CMB}
\end{figure}

In Eq.~(\ref{eq:hydro_approx}), the quantities that depend on the expansion of the universe are integrated over $z$. There are only two such parameters:
$d_A$ and 
$t_L$. 
Their expressions in $\Lambda$CDM and GR-SI are given in Table~\ref{tab:param_mod}. The expressions of the
quantities not explicitly affected by the expansion of the universe  are tabulated in Appendix 
for convenience. Some of these quantities depend indirectly on the expansion of the universe as they contain $t_L$, $z_L$ or $d_A$, namely $d_T$, $R_L$, $d_{\rm Landau}$, $d_{\rm Silk}$, $d_H$ and
$d_D$ (the latter through $d_{\rm Landau}$ and $d_{\rm Silk}$), 
$l_\mathcal{R}$, $l_T$, $l_D$ and $l_H$.
In all, this shows that the Hubble tension may be cast as the problem
of properly modeling the distances $d_A$ and $D_{\mathcal L}$. 
In fact, once SI is accounted for in the CMB anisotropy expression, we can fit the $C^s_{TT,l}$ data 
while keeping $H_0$ to its low-$z$ determination of $73.06$~km/s/Mpc and the $D_M(z)$ parameters obtained from the best fit of $D_{\mathcal L}(z)$ (red line of Fig.~\ref{fig:D(z)}). The parameters allowed to vary are $z_L$, $N$, $n_s$, $\sigma$ and $\Omega_B$, with the $C^s_{TT,l}$ spectrum reproduced 
for 
{ 
${z_L=1728 \pm 1,~N=(1.1995 \pm 0.0019)\times10^{-5},~n_s=0.9759 \pm 0.0028,~
\sigma=1.751 \pm 0.0002}$ and 
${\Omega_Bh^2=0.370 \pm 0.002}$, with ${\chi^2/ndf=0.59}$, see Fig.~\ref{fig:CMB}. We remark
that the quoted uncertainties are only fit uncertainties and do not include other systematic
effects, e.g., coming from approximations in the CMB hydrodynamics model or from the choice of functional form for ${D_M(z)}$ and its parameters.
This fit must use the ${H_0}$ value determined by low-$z$ observations since there is no Hubble tension in the GR-SI model due to the universe expanding differently than in the ${\Lambda}$CDM model. This is verified by performing a CMB fit with ${H_0=67.28}$~km/s/Mpc and observing that the ${\chi^2/ndf}$ of that fit is larger (by about 20\%) than that of the nominal fit. It is also interesting to perform the fit with ${H_0}$ kept a free parameter despite the fact that it introduces a slight inconsistency since the determination of the ${D_M(z)}$ parameters is obtained with the ${H_0}$ value fixed by ${z\simeq 0}$ observations. Such fit yields 
${H_0=72.99\pm0.06}$~km/s/Mpc,
${z_L=1728 \pm 1}$, 
${N=(1.2014 \pm 0.0015)\times10^{-5}}$, 
${n_s=0.9738 \pm 0.0027}$,
${\sigma=1.751 \pm 0.002}$ and 
${\Omega_Bh^2=0.368 \pm 0.002}$, with ${\chi^2/ndf=0.58}$.
}
%

\begin{table}[t]
\caption{CMB quantities depending explicitly on the expansion of the universe. 
Column 1: quantity.
Column 2: $\Lambda$CDM expression.
Column 3: GR-SI expression.
}
\label{tab:param_mod}
\resizebox{1.\textwidth}{!}{%
\begin{tabular}{|c|c|c|} 
\hline
   &     $\Lambda$CDM        &    GR-SI         \\  \hline \hline 
$d_A$  & $\frac{1}{\sqrt{\Omega_K}H_0(1+z_L)}\sinh\left(\sqrt{\Omega_K} \int^1_{1/(1+z_L)} \frac{dx}{\sqrt{\Omega_\Lambda x^4+\Omega_K x^2 +\Omega_M x}}\right)$  &
$\frac{1}{\sqrt{\Omega_K}H_0(1+z_L)}\sinh\left(\sqrt{\Omega_K} \int^1_{1/(1+z_L)} \frac{dx}{\sqrt{\Omega_K x^2 +D_M(1/x-1) x}}\right)$  \\ \hline 	
$t_L$  & $\frac{1}{H_0} \int_0^{1/(1+z_L)} \frac{xdx}{\sqrt{\Omega_\Lambda x^4+\Omega_K x^2 +\Omega_M x +\Omega_R}}$  &
$\frac{1}{H_0} \int_0^{1/(1+z_L)} \frac{xdx}{\sqrt{\Omega_K x^2 +D_M(1/x-1)+\Omega_R]}}$  \\ \hline 	
\end{tabular}}
\end{table}

\section{Conclusion} \label{conclusion}
Our results show that the Hubble tension may be resolved if one accounts, when quantifying the evolution of the universe, for the self-interaction of gravitational fields, a feature of General Relativity ordinarily neglected.
In the cosmological model used in this article, as in the previous studies using that model, the effects of self-interaction are contained within a depletion function which effectively relaxes the traditional assumptions of the Cosmological Principle---isotropy and homogeneity of the evolving universe.
Here, the parameters of the depletion function are determined from the best fit to the luminosity distance data, a procedure that appears more accurate than the method used in~\cite{Deur:2017aas}, {\it viz}, determining the parameters from our knowledge of the timescale at which large-scale structures form, and of the amount of baryonic matter present in these structures. We show that the resulting luminosity distance fits simultaneously both low-redshift supernovae data as well as high-redshift CMB data. 
Furthermore, the model, with the depletion function thus determined, {{ fits better}} the CMB power spectrum data {{ with the ${H_0}$ value determined by the low-${z}$ observations, supporting the finding that there is no Hubble tension in the GR-SI model}}.
Crucially, this possible solution to the problem of the Hubble tension does not require adding parameters beyond those already present in the model. This is important because to be compelling alternate to $\Lambda$CDM, a model should display a consistency and simplicity on par with $\Lambda$CDM, i.e., it should avoid introducing too many new and ad-hoc parameters, particles or fields. This is the case for the model used here which requires no new physics beyond the standard model of particle physics and General Relativity. Explaining the Hubble tension did not compromise this attractive feature of the model.

\section*{Acknowledgements}

This work is done in part with the support of the U.~S.~National Science 
Foundation award No.~1847771. 
The authors are grateful to A. Mand and T. M\"oller for their useful comments on the manuscript.

\begin{table}[h]
\caption{Expressions of the quantities that are not explicitly dependent on the expansion of the universe. 
Column 1: Quantity.
Column 2: $\Lambda$CDM expression.
Column 3: GR-SI expression.
In these expressions, 
$\sigma$ is the standard deviation for the temperature $T_L$, 
$Y\simeq0.24$ is the density ratio of nucleons to neutral $^4$He, 
$\sigma_\mathcal{T}$ is the Thompson cross-section, 
$\rho_B$ and $\rho_\gamma$ are average absolute densities of baryon and radiation, respectively, and
$n_{B0}$ is the baryon number density at present time. }
\label{tab:param_mod 2}
\resizebox{1.\textwidth}{!}{%
\begin{tabular}{|c|c|c|} 
\hline
   &   FLRW universe          &    Universe with GR's SI accounted for         \\  \hline \hline 

$d_T$   & $\sqrt{\Omega_R}/[(1+z_L)H_0\Omega_M]$  &     $\sqrt{\Omega_R}/[(1+z_L)H_0 D_M(0)]$        \\ \hline 	

$R_L$ &  $[3\Omega_B]/[4\Omega_\gamma(1+z_L)]$  &      Same as for $\Lambda$CDM      \\ \hline 	

$R_{EQ}$ & $[3\Omega_R \Omega_B ]/[4\Omega_M \Omega_\gamma]$ & $[3\Omega_R\Omega_B]/[4\Omega_M D(0)\Omega_\gamma]$      \\ \hline

$d_H$  &  $\frac{2}{H_0(3R_L \Omega_M)^{\sfrac{1}{2}}(1+z_L)^{\sfrac{3}{2}}} \ln([\sqrt{1\hspace{-1mm}+\hspace{-1mm}R_L}\hspace{-1mm}+\hspace{-1mm}\sqrt{R_{EQ}\hspace{-1mm}+\hspace{-1mm}R_L}]/[1\hspace{-1mm}+\hspace{-1mm}\sqrt{R_{EQ}}])$
  & $\frac{2}{H_0(3R_L D_M(0))^{\sfrac{1}{2}}(1+z_L)^{\sfrac{3}{2}}} \ln([\sqrt{1\hspace{-1mm}+\hspace{-1mm}R_L}\hspace{-1mm}+\hspace{-1mm}\sqrt{R_{EQ}\hspace{-1mm}+\hspace{-1mm}R_L}]/[1\hspace{-1mm}+\hspace{-1mm}\sqrt{R_{EQ}}])$       \\ \hline 	

$d_D$ & $\sqrt{d^2_{\rm Landau}+d^2 _{\rm Silk}}$ & Same as for $\Lambda$CDM          \\ \hline 	

$d^2 _{\rm Landau}$ & $\frac{3\sigma^2 t_L^2}{8T_L^2 (1+R_L)},$ & Same as for $\Lambda$CDM          \\ \hline 	

$d^2 _{\rm Silk}$   & $\frac{R_L^2}{6(1-Y)(n_{B0})\sigma_\mathcal{T}H_0 \sqrt{\Omega_M}R_0^{\sfrac{9}{2}}} 
\int_0^{R_L} \frac{R^2 dR}{X(R)(1+R)\sqrt{R_{EQ}+R}}\big[\frac{16}{15}\hspace{-1mm}+\hspace{-1mm}\frac{R^2}{1+R}\big]$  &       
 $\frac{R_L^2}{6(1-Y)(n_{B0})\sigma_\mathcal{T}H_0 \sqrt{D_M(0)}R_0^{\sfrac{9}{2}}} 
\int_0^{R_L} \frac{R^2 dR}{X(R)(1+R)\sqrt{R_{EQ}+R}}\big[\frac{16}{15}\hspace{-1mm}+\hspace{-1mm}\frac{R^2}{1+R}\big]$   \\ \hline %

$R(t)$ & $ \sfrac{3\rho_B(t)}{4\rho_\gamma(t)}$& Same as for $\Lambda$CDM          \\ \hline

$X(T)$  &   $1/\big[ X^{-1}(3400)+\frac{\Omega_B h^2}{(\Omega_M h^2)^{1/2}} \int_T^{3400} g(T')dT' \big]$ &       
$1/\big[ X^{-1}(3400)+\frac{\Omega_B h^2}{(D_M(0) h^2)^{1/2}} \int_T^{3400} g(T')dT' \big]$     \\ \hline 	
\end{tabular}}
\end{table}
%



\begin{references}


\bibitem{Planck:2018vyg}
N.~Aghanim \textit{et al.} [Planck],
Astron. Astrophys. \textbf{641} (2020), A6
[erratum: Astron. Astrophys. \textbf{652} (2021), C4]
[arXiv:1807.06209 [astro-ph.CO]].

\bibitem{eBOSS:2020yzd}
S.~Alam \textit{et al.} [eBOSS],
Phys. Rev. D \textbf{103} (2021) no.8, 083533
[arXiv:2007.08991 [astro-ph.CO]].

\bibitem{Riess:2016jrr}
A.~G.~Riess, L.~M.~Macri, S.~L.~Hoffmann, D.~Scolnic, S.~Casertano, A.~V.~Filippenko, B.~E.~Tucker, M.~J.~Reid, D.~O.~Jones and J.~M.~Silverman, \textit{et al.}
Astrophys. J. \textbf{826} (2016) no.1, 56
[arXiv:1604.01424 [astro-ph.CO]].

\bibitem{Wong:2019kwg}
K.~C.~Wong, S.~H.~Suyu, G.~C.~F.~Chen, C.~E.~Rusu, M.~Millon, D.~Sluse, V.~Bonvin, C.~D.~Fassnacht, S.~Taubenberger and M.~W.~Auger, \textit{et al.}
Mon. Not. Roy. Astron. Soc. \textbf{498} (2020) no.1, 1420-1439
[arXiv:1907.04869 [astro-ph.CO]].

\bibitem{Abdalla:2022yfr}
E.~Abdalla, G.~Franco Abell\'an, A.~Aboubrahim, A.~Agnello, O.~Akarsu, Y.~Akrami, G.~Alestas, D.~Aloni, L.~Amendola and L.~A.~Anchordoqui, \textit{et al.}
JHEAp \textbf{34} (2022), 49-211
[arXiv:2203.06142 [astro-ph.CO]].


\bibitem{Verde:2019ivm}
L.~Verde, T.~Treu and A.~G.~Riess,
Nature Astron. \textbf{3}, 891
[arXiv:1907.10625 [astro-ph.CO]].

\bibitem{DiValentino:2021izs}
E.~Di Valentino, O.~Mena, S.~Pan, L.~Visinelli, W.~Yang, A.~Melchiorri, D.~F.~Mota, A.~G.~Riess and J.~Silk,
Class. Quant. Grav. \textbf{38} (2021) no.15, 153001
[arXiv:2103.01183 [astro-ph.CO]].

\bibitem{Riess_2022}
A.~G.~Riess, W.~Yuan, L.~M.~Macri, D.~Scolnic, D.~Brout, S.~Casertano, D.~O.~Jones, Y.~Murakami, L.~Breuval and T.~G.~Brink, \textit{et al.}
Astrophys. J. Lett. \textbf{934} (2022) no.1, L7
[arXiv:2112.04510 [astro-ph.CO]].

\bibitem{Freedman_2020}
W.~L.~Freedman, B.~F.~Madore, T.~Hoyt, I.~S.~Jang, R.~Beaton, M.~G.~Lee, A.~Monson, J.~Neeley and J.~Rich,
[arXiv:2002.01550 [astro-ph.GA]].

\bibitem{Kahlhoefer:2017dnp}
F.~Kahlhoefer,
Int. J. Mod. Phys. A \textbf{32} (2017) no.13, 1730006
[arXiv:1702.02430 [hep-ph]].

\bibitem{Gaskins:2016cha}
J.~M.~Gaskins,
Contemp. Phys. \textbf{57} (2016) no.4, 496-525
[arXiv:1604.00014 [astro-ph.HE]].

\bibitem{Arcadi:2017kky}
G.~Arcadi, M.~Dutra, P.~Ghosh, M.~Lindner, Y.~Mambrini, M.~Pierre, S.~Profumo and F.~S.~Queiroz,
Eur. Phys. J. C \textbf{78} (2018) no.3, 203
[arXiv:1703.07364 [hep-ph]].

\bibitem{Klypin:1999uc}
A.~A.~Klypin, A.~V.~Kravtsov, O.~Valenzuela and F.~Prada,
Astrophys. J. \textbf{522} (1999), 82-92
[arXiv:astro-ph/9901240 [astro-ph]].

\bibitem{Tully:1977fu}
R.~B.~Tully and J.~R.~Fisher,
Astron. Astrophys. \textbf{54} (1977), 661-673

\bibitem{McGaugh:2016leg}
S.~McGaugh, F.~Lelli and J.~Schombert,
Phys. Rev. Lett. \textbf{117} (2016) no.20, 201101
[arXiv:1609.05917 [astro-ph.GA]].

\bibitem{renzo}
R.~Sancisi,
IAU Symp. \textbf{220} (2004), 233
[arXiv:astro-ph/0311348 [astro-ph]].

\bibitem{Deur:2009ya}
A.~Deur,
Phys. Lett. B \textbf{676} (2009), 21-24
[arXiv:0901.4005 [astro-ph.CO]].

\bibitem{Deur:2020wlg}
A.~Deur,
Eur. Phys. J. C \textbf{81} (2021) no.3, 213
[arXiv:2004.05905 [astro-ph.GA]].

\bibitem{Deur:2017aas}
A.~Deur,
Eur. Phys. J. C \textbf{79} (2019) no.10, 883
[arXiv:1709.02481 [astro-ph.CO]].

\bibitem{Deur:2022ooc}
A.~Deur,
Class. Quant. Grav. \textbf{39} (2022) no.13, 135003
[arXiv:2203.02350 [gr-qc]].

\bibitem{Deur:2021ink}
A.~Deur,
Phys. Lett. B \textbf{820} (2021), 136510
[arXiv:2108.04649 [physics.gen-ph]].

\bibitem{Deur:2019kqi}
A.~Deur, C.~Sargent and B.~Terzi\'c,
Astrophys. J. \textbf{896} (2020) no.2, 94
[arXiv:1909.00095 [astro-ph.GA]].

\bibitem{Zee}
A.~Zee,
Princeton University Press, 2013,
ISBN 978-0-691-14558-7


\bibitem{Fierz:1939ix}
M.~Fierz and W.~Pauli,
Proc. Roy. Soc. Lond. A \textbf{173} (1939), 211-232

\bibitem{Deur:2016tte}
A.~Deur, S.~J.~Brodsky and G.~F.~de Teramond,
Nucl. Phys. \textbf{90} (2016), 1
[arXiv:1604.08082 [hep-ph]].

\bibitem{Deur:2013baa}
A.~Deur,
Mon. Not. Roy. Astron. Soc. \textbf{438} (2014) no.2, 1535-1551
[arXiv:1304.6932 [astro-ph.GA]].

\bibitem{Winters:2022tew}
D.~Winters, A.~Deur and X.~Zheng,
Mon. Not. Roy. Astron. Soc. \textbf{518} (2022) no.2, 2845-2852
[arXiv:2207.02945 [astro-ph.GA]].

\bibitem{Deur_DM-EPJC}
A.~Deur,
Eur. Phys. J. C \textbf{77} (2017) no.6, 412
[arXiv:1611.05515 [hep-ph]].


\bibitem{Weinberg:2008zzc}
S.~Weinberg,

\bibitem{Aghanim:2019ame}
N.~Aghanim \textit{et al.} [Planck],
Astron. Astrophys. \textbf{641} (2020), A5
[arXiv:1907.12875 [astro-ph.CO]].



\end{references}
\end{document}